\documentclass[prc,twocolumn,aps,superscriptaddress,nofootinbib]{revtex4-1}
\usepackage{amssymb}
\usepackage{amsmath,bm}
\usepackage{graphicx}
\usepackage[normalem]{ulem}
\usepackage{bbding}
\usepackage{booktabs}
\usepackage[usenames, dvipsnames]{xcolor}
\usepackage{lipsum}
\usepackage{multirow}
\usepackage{multirow}

\usepackage{makecell}
\usepackage{colortbl}

\usepackage{threeparttable}

\usepackage[colorlinks,
bookmarks=true,
linkcolor=blue,
urlcolor=blue,
anchorcolor=black,
citecolor=blue
]{hyperref}

\renewcommand{\sout}{\bgroup \color{red} \ULdepth=-.5ex \ULset}

\usepackage{isotope}

\AtBeginDocument{
\heavyrulewidth=.08em
\lightrulewidth=.05em
\cmidrulewidth=.03em
\belowrulesep=.65ex
\belowbottomsep=0pt
\aboverulesep=.4ex
\abovetopsep=0.2pt
\cmidrulesep=0.5pt
\cmidrulekern=.5em
\defaultaddspace=2.5em
}
\begin{document}


\title{Bayesian inference of the symmetry energy and the neutron skin in $^{48}$Ca and $^{208}${Pb} from CREX and PREX-2}

\author{Zhen Zhang}
\thanks{Corresponding author}
\email{zhangzh275$@$mail.sysu.edu.cn}
\affiliation{Sino-French Institute of Nuclear Engineering and Technology,
Sun Yat-sen University, Zhuhai 519082, China}

\author{Lie-Wen Chen}
\thanks{Corresponding author}
\email{lwchen$@$sjtu.edu.cn}
\affiliation{School of Physics and Astronomy, Shanghai Key Laboratory for
Particle Physics and Cosmology, and Key Laboratory for Particle Astrophysics and Cosmology (MOE),
Shanghai Jiao Tong University, Shanghai 200240, China }

\date{\today}

\begin{abstract}

Using the recent model-independent determination of the charge-weak form factor difference $\Delta F_{\rm CW}$ in \isotope[48]{Ca} and \isotope[208]{Pb} by the CREX and PREX-2 collaborations  together with some well-determined properties of doubly magic nuclei, we perform Bayesian inference of the symmetry energy $E_{\rm sym}(\rho)$ and the neutron skin thickness $\Delta r_{\rm np}$ of $^{48}$Ca and $^{208}${Pb} within the Skyrme energy density functional~(EDF).
We find
the inferred $E_{\rm sym}(\rho)$ and $\Delta r_{\rm np}$ separately from CREX and PREX-2 are compatible with each other at $90\%$ C.L., although they are inconsistent at $68.3\%$ C.L. with CREX (PREX-2) favoring a very soft (stiff) $E_{\rm sym}(\rho)$ and rather small (large) $\Delta r_{\rm np}$.
By combining the CREX and PREX-2 data, we obtain a soft symmetry energy around saturation density $\rho_0$ and thinner $\Delta r_{\rm np}$ of $^{48}$Ca and $^{208}${Pb}, which are found to be
closer to the corresponding results from CREX alone, implying the PREX-2 is less effective to constrain the $E_{\rm sym}(\rho)$ and $\Delta r_{\rm np}$ due to its lower precision of $\Delta F_{\rm CW}$.
Furthermore,
we find the Skyrme EDF results inferred by combining the CREX and PREX-2 data
nicely agree with the measured dipole polarizabilities $\alpha_D$ in \isotope[48]{Ca} and \isotope[208]{Pb} as well as the neutron matter equation of state from microscopic calculations.
The implications of the inferred soft $E_{\rm sym}(\rho)$ around $\rho_0$ are discussed.

\end{abstract}

\maketitle

\section{Introduction}
The CREX~\cite{CREX:2022kgg} and PREX-2~\cite{PREX:2021umo} collaborations recently reported
the model-independent extractions of the difference between
the charge form factor $F_C$ and the weak form factor $F_{W}$, i.e., $\Delta F_{{\rm{CW}}}(q)\equiv F_C(q) - F_W(q) = 0.0277 \pm 0.0055 $ at $q= 0.8733~\rm{fm}^{-1}$ for \isotope[48]{Ca} and $\Delta F_{\rm{CW}}(q)=0.041\pm0.013$ at a smaller four-momentum transfer $q = 0.3977~\rm{fm}^{-1}$ for \isotope[208]{Pb}~\cite{CREX:2022kgg}.
Since these extractions are free from the strong interaction uncertainties,
they allow to determine with minimal model-dependence the neutron skin thickness $\Delta r_{\rm np}\equiv r_n - r_p$ [$r_{n(p)}$ is the neutron(proton) rms radius of the nucleus] and further to constrain the density dependence of the symmetry energy $E_{\rm sym}(\rho)$~\cite{Brown:2000pd,Typel:2001lcw,Horowitz:2000xj,Furnstahl:2001un,Chen:2005ti,Centelles:2008vu,Warda:2009tc,Roca-Maza:2011qcr,Agrawal:2012pq,Zhang:2013wna}.
The $E_{\rm sym}(\rho)$ encodes the isospin dependence of nuclear matter equation of state~(EOS) and plays an important role in both nuclear physics and astrophysics~\cite{Baran2005,Steiner2005b,Lattimer2007a,Li2008}.

From the PREX-2 data, the $\Delta r_{\rm np}$ of \isotope[208]{Pb} is extracted to be $0.283\pm0.071~\rm{fm}$~\cite{PREX:2021umo}. An analysis based on a relativistic energy density functional~(EDF) indicates the PREX-2 data lead to a very stiff $E_{\rm sym}(\rho)$ with a rather large symmetry energy slope parameter [$L(\rho_r)=3\rho_r \frac{dE_{\rm{sym}}(\rho)}{d\rho}|_{\rho_r}$] of $L\equiv L(\rho_0)=106\pm37$ MeV at saturation density $\rho_0$~\cite{Reed:2021nqk}, which  challenges our present understanding on the $E_{\rm sym}(\rho)$~\cite{Lattimer2014,Li2013a,Oertel:2016bki}.
Many studies have been devoted to understanding the PREX-2 result and its implications in nuclear physics and astrophysics~\cite{Yue:2021yfx,Reed:2021nqk,Piekarewicz:2021jte,Reinhard:2021utv,Biswas:2021yge}.
In particular, the tension between the PREX-2 data and the measured electric dipole polarizabilities $\alpha_D$ in \isotope[48]{Ca} and \isotope[208]{Pb} at RCNP in Osaka~\cite{Tamii2011a,Roca-Maza:2015eza,Birkhan:2016qkr} is observed with the latter favoring a much softer $E_{\rm sym}(\rho)$~\cite{Piekarewicz:2021jte,Reinhard:2021utv}.

Very remarkably, the CREX adopts the same experimental approach as PREX-2 and recently report a rather thin neutron skin of $\Delta r_{\rm np}=0.121\pm0.026(\rm exp)\pm 0.024(\rm model)$~fm in \isotope[48]{Ca}~\cite{CREX:2022kgg}. Analyses with a number of modern nonrelativistic and relativistic EDFs~\cite{Reinhard:2022inh,Yuksel:2022umn} (see also Ref.~\cite{CREX:2022kgg}) suggest a significant tension between the CREX and PREX-2 results, calling for further critical theoretical and experimental investigations.

In this work, we employ the Bayesian inference method, which provides a consistent probabilistic approach to
extract quantitative information from experimental data~\cite{RevModPhys.83.943},
to analyze the CREX and PREX results on the $\Delta F_{{\rm{CW}}}(q)$ together with other well-known data of eight doubly magic nuclei, i.e., \isotope[16]{O}, \isotope[40]{Ca}, \isotope[48]{Ca}, \isotope[56]{Ni}, \isotope[68]{Ni}, \isotope[100]{Sn}, \isotope[132]{Sn} and \isotope[208]{Pb}, based on the Skyrme EDF.
We show the CREX and PREX-2 results are compatible at $90\%$ confidence level~(C.L.), although they are inconsistent with each other at $68.3\%$ C.L., and furthermore the PREX-2 is less effective to constrain the $E_{\rm sym}(\rho)$ and $\Delta r_{\rm np}$ due to its lower precision of $\Delta F_{\rm CW}$ compared to the CREX.
By combining the CREX and PREX-2 results at $90\%$ C.L.,
we find a soft $E_{\rm sym}(\rho)$ around $\rho_0$ can be inferred and the Skyrme EDF can nicely describe the measured $\alpha_D$ in \isotope[48]{Ca} and \isotope[208]{Pb} as well as the neutron matter EOS from microscopic many-body calculations.

\section{Model and method}
The nuclear properties are calculated within the widely used standard Skyrme EDF.
Since we focus on doubly magic nuclei, pairing interaction is not taken
into account. The Skyrme EDF can then be characterized by ten parameters: the $\rho_0$, the binding energy per nucleon of symmetric nuclear matter $E_0(\rho_0)$, the incompressibility $K_0$, $E_{\mathrm{sym}}(\rho_0)$, $L$, the isoscalar effective mass $m_{s,0}^*$ and the isovector effective mass $m_{v,0}^*$ at $\rho_0$, the gradient coefficient $G_S$, the symmetry-gradient coefficient $G_{V}$, and the spin-orbit coupling constant $W_0$~\cite{Chen2010,Chen:2011ps,Kortelainen2010}.
Based on the Skyrme EDF, once given a parameter set
\begin{eqnarray}
\bm{p} & =& \lbrace  \rho_0,~E_0(\rho_0),~K_0,~E_{\mathrm{sym}}(\rho_0),~L, \notag \\
& &~ G_S,~G_V,~W_0,~m_{s,0}^{\ast},~m_{v,0}^{\ast}\rbrace,
\end{eqnarray}
the ground-state properties of finite nuclei are calculated with the Hartree-Fock~(HF) method, and the breathing mode energy is obtained from the constrained HF~(CHF) calculation.

\begin{table}[htb]
\caption{Prior ranges of the ten parameters used, together with the posterior median values and $68.3\%$($90\%$) credible intervals from B-All and the parameter values of the Skyrme interaction SkREx.}
\label{Tab:Prior}
\begin{tabular}{lccc}
\hline \hline
Quantity & prior &  posterior~(B-All) & SkREx\\
\hline
$\rho_0~(\mathrm{fm}^{-3})$                       &     $[  0.155 ,0.165]$    & $0.1619_{-0.0016(0.0026)}^{+0.0015(0.0023)}$       &0.1618       \\
$E_0~(\mathrm{MeV}) $                             &     $[-16.5   ,-15.5]$    & $-16.002_{-0.063(0.103)}^{+0.063(0.100)}$     &-16.00   \\
$K_0~(\mathrm{MeV})$                              &     $[ 210    , 250]$     & $225.0_{-2.8(4.6)}^{+2.9(4.9)}$     &223.1 \\
$E_{\mathrm{sym}}(\rho_0)~(\mathrm{MeV})$         &     $[ 22     ,  55 ]$    & $29.1_{-1.8(2.7)}^{+2.1(3.6)}$      &29.2 \\
$L~(\mathrm{MeV})$                                &     $[ -90    ,  240]$    & $17.1_{-22.3(36.0)}^{+23.8(39.3)}$  &13.0 \\
$G_S~(\mathrm{MeV}\cdot \mathrm{fm}^{5})$          &    $[ 110    ,  170]$    & $117.9_{-4.2(6.2)}^{+6.4(11.8)}$    &118.9 \\
$G_V~(\mathrm{MeV}\cdot \mathrm{fm}^{5})$          &    $[ -70    ,   70]$    & $-27.3_{-31.0(39.1)}^{+46.5(73.6)}$ &-55.0 \\
$W_0~(\mathrm{MeV}\cdot \mathrm{fm}^{5})$          &    $[ 90     ,  140]$    & $105.4_{-4.9(7.9)}^{+5.0(8.3)}$     &117.2 \\
$m_{s,0}^{\ast}/m$                                &     $[  0.7   ,  1.0]$    & $0.95_{-0.06(0.10)}^{+0.03(0.04)}$       &0.969    \\
$m_{v,0}^{\ast}/m$                                &     $[  0.6   ,  0.9]$    & $0.71_{-0.08(0.10)}^{+0.11(0.16)}$       &0.640   \\
\hline\hline
\end{tabular}
\end{table}

The calibration and uncertainty quantification of the ten parameters is carried out using a Bayesian approach. According to Bayes' theorem, the posterior distribution of model parameters $\bm{p}$, given experimental data $\mathcal{O}^{\rm{exp} }$ for a set of observables $\mathcal{O}$, can be evaluated as
\begin{eqnarray}
\label{Eq:Baye}
P\left(\boldsymbol{p} \mid \mathcal{M}, \mathcal{O}^{\exp }\right)=\frac{P\left( \mathcal{O}^{\exp } \mid \mathcal{M},\boldsymbol{p}\right) P(\boldsymbol{p})}{\int P\left(\mathcal{O}^{\exp } \mid \mathcal{M}, \boldsymbol{p}\right) P(\boldsymbol{p}) \mathrm{d} \boldsymbol{p}},
\end{eqnarray}
where $\mathcal{M}$ is the given model, $P(\boldsymbol{p})$ is the prior probability density of model parameters $\boldsymbol p$ before being confronted with the data $\mathcal{O}^{\exp }$, and $P\left( O^{\exp } \mid \mathcal{M}, \boldsymbol{p}\right)$ denotes the likelihood of observing $\mathcal{O}^{\exp }$ with given model $\mathcal{M}$ predictions at $\boldsymbol p$.
The prior distribution of $\bm{p}$ are normally chosen to be uniform in their empirical
ranges listed in Tab.~\ref{Tab:Prior}. In particular, given the rather thick (thin) neutron skin in \isotope[208]{Pb}~(\isotope[48]{Ca}) from PREX-2~(CREX), the prior ranges of $E_{\mathrm{sym}}(\rho_0)$ and $L$ are taken to be as large as $22\sim 55$ MeV and $-90\sim 240$ MeV, respectively, to avoid the prior range dependence of the posterior results.

The likelihood function is taken to be the commonly used Gaussian form
\begin{eqnarray}
\label{Eq:likelihood}
P\left(\mathcal{M}, O_{i}^{\exp } \mid \boldsymbol{p}\right) \propto \exp \left\{-\sum_{i} \frac{\left[\mathcal{O}_{i}(\boldsymbol{p})-\mathcal{O}_{i}^{\exp }\right]^{2}}{2 \sigma_{i}^{2}}\right\},
\end{eqnarray}
where $\mathcal{O}_{i}(\boldsymbol{p})$ is the model prediction on the $i$-th observable for a given parameter set $\boldsymbol{p}, \mathcal{O}_{i}^{\exp }$ is the corresponding data, and $\sigma_{i}$ is the adopted error.

To estimate the posterior distribution given by  Eq.~(\ref{Eq:Baye}), the Markov chain Monte Carlo (MCMC) process is carried out using the Metropolis-Hasting algorithm. We  first run $5\times10^5$ burn-in MCMC steps to allow the chain to reach equilibrium,  and then generate $10^6$ MCMC steps in parameter space. The posterior distributions of model parameters and observables are estimated from 15 parallel MCMC process, i.e., $1.5\times 10^7$ MCMC samples.

  \begin{table}[htbp]
      \caption{Experimental data and adopted errors used in the Bayesian analysis. The second line shows the globally adopted error for each observable. That error is multiplied for each observable by a further integer weight factor given in the parenthesis next to the data value.
      For the data and adopted errors of the neutron-proton Fermi energy differences $\Delta \epsilon_F$ of \isotope[16]{O}, \isotope[40]{Ca}, \isotope[48]{Ca}, \isotope[56]{Ni}, \isotope[132]{Sn} and \isotope[208]{Pb} as well as the breathing mode energy $E_{\rm{GMR}}$ of \isotope[208]{Pb}, see the text. 
      For the charge-weak form factor difference
      $\Delta F_{{\rm{CW}}}$,
      the CREX and PREX-2 results, i.e., $\Delta F_{{\rm{CW}}}(q= 0.8733~\rm{fm}^{-1}) = 0.0277 \pm 0.0055$ for \isotope[48]{Ca} and $\Delta F_{\rm{CW}}(q = 0.3977~\rm{fm}^{-1})=0.041\pm0.013$ for \isotope[208]{Pb}~\cite{CREX:2022kgg}, are used in this work.
      }
      \label{Tab:Data}
      \begin{threeparttable}
      \begin{tabular}{c*{5}{c}}
        \hline\hline
      Nuclei
      & {$E_{\rm{B}}$}
      & {$r_c$}
      & {$R_{{d}}$}
      & {$\sigma$}
      & {$\Delta\epsilon_{ls}$} \\
      & {($1~\rm{MeV}$) }
      & {($0.02~\rm{fm}$)}
      & {($0.04~\rm{fm}$)}
      & {($0.04~\rm{fm}$)}
      & {($ 20\%$)}         \\
      \hline
       ${}^{16}$O      & $-127.620 (4)$  & $2.701(2)$ & $2.777(2)$ & $0.839(2)$  & $6.30(3)$\\
                            &                 & $      $   & $      $ & $     $       & $6.10(3)$\\
       ${}^{40}$Ca     & $-342.051 (3)$  & $3.478(1)$ & $3.845(1)$ & $0.978(1)$  &                      \\
       ${}^{48}$Ca     & $-415.990 (1)$  & $3.479(2)$ & $3.964(1)$ & $0.881(1)$  &                      \\
       ${}^{56}$Ni     & $-483.990 (5)$  & $3.750(9)$ &            &             &                      \\
       ${}^{68}$Ni     & $-590.430 (1)$  &            &            &             &                      \\
      ${}^{100}$Sn   & $-825.800 (2)$  &            &            &             &                      \\
      ${}^{132}$Sn   & $-1102.900(1)$  &            &            &             & $1.35(1)$ \\
                           &                 &            &            &             & $1.65(1)$  \\
      ${}^{208}$Pb   & $-1636.446(1)$  & $5.504(1)$ & $6.776(1)$ & $0.913(1)$  & $1.32(1)$ \\
                           &                   &             &                      &         & $0.90(1)$ \\
                           &                   &             &                      &         & $1.77(2)$\\
      \hline\hline
                          \end{tabular}
     \begin{tablenotes}
     \item[] {\bf Note. }$\Delta \epsilon_{ls}$ data are for  $^{16}$O($1p_p$, $1p_n$), $^{132}$Sn($2p_p$, $2d_n$),
     and  $^{208}$Pb($2d_p$, $3p_n$, $2f_n$), respectively.
     \end{tablenotes}
   \end{threeparttable}
      \end{table}

\section{\label{Sec:Ob}Selected observables}
The key observables in this work are the model-independent $\Delta F_{\rm{CW}}$ in \isotope[48]{Ca} and \isotope[208]{Pb}, i.e.,
$\Delta F_{\rm{CW}}^{48}\equiv \Delta F_{\rm{CW}}(q=0.8733~\rm{fm}^{-1})$ from CREX~\cite{CREX:2022kgg} and $\Delta F_{\rm{CW}}^{208} \equiv \Delta F_{\rm{CW}}(q=0.3977~\rm{fm}^{-1})$ from PREX-2~\cite{PREX:2021umo}.
The normalized nuclear form factors $F_{C}(q)$ and $F_{W}(q)$ are calculated by folding the nucleon form factor $F_t(q)~(t=n,p)$ and the spin-orbit current form factor ($F_t^{ls}$) with the intrinsic nucleon electromagnetic form factor $G_{E/M,t}$ and weak form factor $G^{(W)}_{E/M,t}$ by~\cite{Reinhard:2021utv}
\begin{eqnarray}
F_{C}(q) &=&\frac{1}{Z} \sum_{t=p, n}\left[G_{E, t}(q) F_{t}(q)+G_{M, t}(q) F_{t}^{(l s)}(q)\right],\\
F_{W}(q) &=&\sum_{t=p, n}\frac{\left[G_{E, t}^{(W)}(q) F_{t}(q)+G_{M, t}^{(W)}(q) F_{t}^{(l s)}(q)\right]}{Z Q_{p}^{(W)}+N Q_{n}^{(W)}},
\end{eqnarray}
where $N(Z)$ is the neutron(proton) number, and  $Q_p^{(W)} = 0.0713$ and $Q_n^{(W)}=-0.9888$ are proton and neutron weak charges, respectively.
The $G_{E/M,t}$ are derived from the isospin-coupled Sachs form factors, the relativistic Darwin correction has been included and the center-of-mass corrections are taken into account by simply renormalizing the nucleon mass $m_N$ to $(1-1/A)m_N$ in the HF calculation (see, e.g., Ref.~\cite{Bender:2003jk} for details).
From $G_{E/M,t}$, the $G^{(W)}_{E/M,t}$ can then be determined by further considering the contribution of the strange-quark electromagnetic form factors $G_{E/M,s}$ (see, Ref.~\cite{Reinhard:2021utv} for details).
In addition,
the $F_C(q)$ at low momentum $q$ can be characterized
by three parameters~\cite{Friedrich:1982esq}, i.e., the charge rms radius
\begin{equation}
r_c = \sqrt{-\left.\frac{3}{F_C(0)}\frac{d^2}{dq^2}F_C(q)\right\vert_{q=0}},
\end{equation}
the diffraction radius $R_d=\frac{4.493}{q_0}$ determined from the first zero of $F_C[q_0]=0$, and the surface
thickness
\begin{equation}
\sigma= \sqrt{\frac{2}{q_{m}} \log \left[\frac{3j_1(q_mR _d)}{q_m R_d F_{C}\left(q_{m}\right)}\right]}, \quad q_{m}=5.6 / R,
\label{Eq:sg}
\end{equation}
where $j_1(x) = \frac{\sin x}{x^2}-\frac{\cos x}{x}$ is the spherical Bessel function of the first kind.

In this work, we also
include in our analysis some well-determined data, i.e., the total binding energies $E_{{B}}$, $r_c$, $R_{{d}}$,
$\sigma$, spin-orbit splittings $\Delta\epsilon_{ls}$, neutron-proton Fermi energy differences $\Delta \epsilon_F$, and breathing mode energies $E_{\rm GMR}$ of doubly magic nuclei: \isotope[16]{O}, \isotope[40]{Ca}, \isotope[48]{Ca}, \isotope[56]{Ni}, \isotope[68]{Ni}, \isotope[100]{Sn}, \isotope[132]{Sn}, \isotope[208]{Pb}.
The values and adopted errors for $E_B$, $r_c$, $R_d$,
$\sigma$ and $\Delta \epsilon_{ls}$ are taken from Ref~\cite{Klupfel:2008af}, and listed in Table~\ref{Tab:Data} for completeness. For $\Delta \epsilon_F$, the experimental values for \isotope[16]{O}, \isotope[40]{Ca}, \isotope[48]{Ca}, \isotope[56]{Ni},  \isotope[132]{Sn}, and \isotope[208]{Pb} are $-3.53$, $-7.31$, $6.10$, $-9.47$, $8.40$ and $0.64$ MeV, respectively~\cite{Wang2013}. According to calculations for $\Delta \epsilon_F$ of 19 nuclei with 54 Skyrme interactions reported in Ref.~\cite{Wang2013}, we take their uncertainties to be $1.2$ MeV.
As to the $E_{\rm{GMR}}$ of \isotope[208]{Pb}, the weighted average\footnote{Here, the weighted average  is defined in the standard way, i.e., given $n$ independent measured value $\mathcal{O}_i$  with the standard deviation $\sigma_i$ for the same observable,  the weighted average  is calculated as
$\bar{\mathcal{O}} = \frac{\sum_{i=1}^n \mathcal{O}_i/\sigma_i}{\sum_{i=1}^n \sigma_i^{-1}},$ and its
standard deviation  is $\sigma_{\overline{\mathcal{O}}}=\left(\sum_{i=1}^n \sigma_i^{-2}\right)^{-1 / 2}$.}
$13.614\pm0.074$ MeV of two independently measured values by RCNP~\cite{Patel2013,Howard:2020pnz} and TAMU~\cite{Youngblood:1999zza} is used.

\section{Results and discussions.}
Bayesian analyses are conducted in four different cases in the present work. The case without including the $\Delta F_{\rm{CW}}^{208}$ and $\Delta F_{\rm{CW}}^{48}$ is considered as the base point labeled with ``B-Bas".
The $\Delta F_{\rm{CW}}^{48}$ and $\Delta F_{\rm{CW}}^{208}$ data are then separately
added into the analysis to quantify the tension between the CREX and PREX data, and the results are accordingly labeled by ``B-$\Delta F_{\rm{CW}}^{48}$" and ``B-$\Delta F_{\rm{CW}}^{208}$".
A Bayesian analysis including all the selected observables  labeled by ``B-All'' is further carried out to constrain the $E_{\rm sym}(\rho)$ by combining the CREX and PREX results. The posterior median values and $68.3\%(90\%)$ credible intervals of parameters $\bm{p}$
obtained with B-All are listed in Tab.~\ref{Tab:Prior}.

\begin{figure}[htb]
  \centering
  \includegraphics[width=0.95\linewidth]{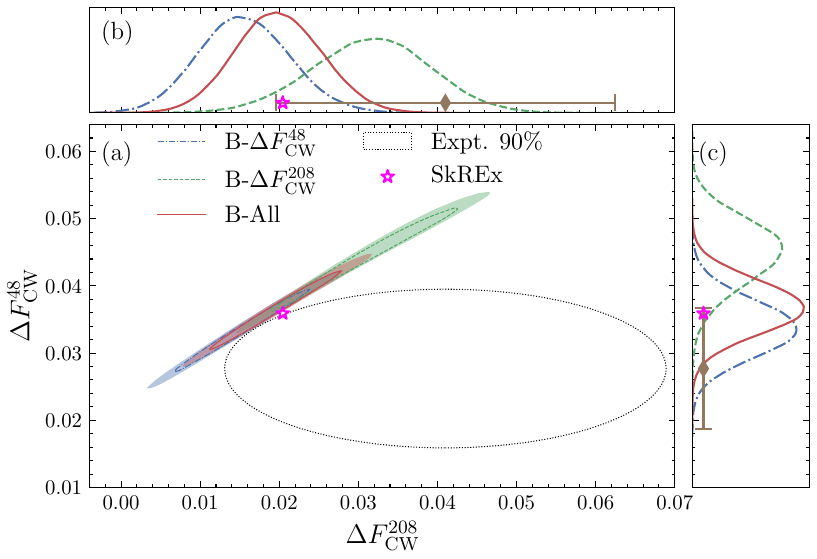}
  \caption{(Color online). Joint [(a)] and marginal [(b) and (c)] distributions of  {$\Delta F_{\rm{CW}}^{208}$ and $\Delta F_{\rm{CW}}^{48}$} from B-$\Delta F_{\rm{CW}}^{48}$ (blue, dash-dotted line), B-$\Delta F_{\rm{CW}}^{208}$ (green, dashed line) and B-All (red, solid line). The shaded regions and the lines correspond to
  $68.3\%$ and $90\%$ credible regions, respectively. The $90\%$ credible regions of the experimental joint and marginal distributions by CREX and PREX-2 are indicated as the dotted ellipse and solid diamonds, respectively.}
\label{Fig:DFCW}
\end{figure}

Figure~\ref{Fig:DFCW} shows the obtained posterior joint [(a)] and marginal ([(b) and (c)]) distributions of $\Delta F_{\rm{CW}}^{208}$ and $\Delta F_{\rm{CW}}^{48}$ from B-$\Delta F_{\rm{CW}}^{48}$, B-$\Delta F_{\rm{CW}}^{208}$ and B-All,
together with
the corresponding experimental joint distribution of $90\%$ credible region as well as the CREX and PREX individual measurements with $90\%$ uncertainties.
It is seen that due to the constraints from properties of doubly magic nuclei, the inferred $\Delta F_{\rm{CW}}^{208(48)}$ from B-$\Delta F_{\rm{CW}}^{208(48)}$ is less (larger) than the PREX-2 (CREX) measurement. In all the three cases, the inferred $90\%$ confidence regions
barely overlap with the experimental one, which indicates the tension between the CREX and PREX-2 results within the framework of Skryme EDF.
From the MCMC samples,
we find out a Skyrme interaction (named ``SkREx'') that is consistent with both the CREX and PREX data at $90\%$ C.L. (as indicated by star in Fig.~\ref{Fig:DFCW}), the measured $\alpha_D$ in \isotope[48]{Ca} and \isotope[208]{Pb}, and the neutron matter EOS from microscopic calculations as shown later. The parameter values of SkREx are listed in the last column of Tab.~\ref{Tab:Prior}.
The 9 Skyrme parameters of SkREx are: $t_0= -2088.20~\rm{MeV~fm}^3$, $x_0=0.285971$, $t_1=322.498~\rm{MeV~fm}^5$, $x_1=0.760722$, $t_2 = 537.638~\rm{MeV~fm}^5$, $x_2 = -1.66900$, $t_3 =13965.6~\rm{MeV~fm}^{3+3\alpha}$, $x_3 =0.0165947$, and $\alpha = 0.261515$.

\begin{figure}[bth]
    \centering
    \includegraphics[width=0.9\linewidth]{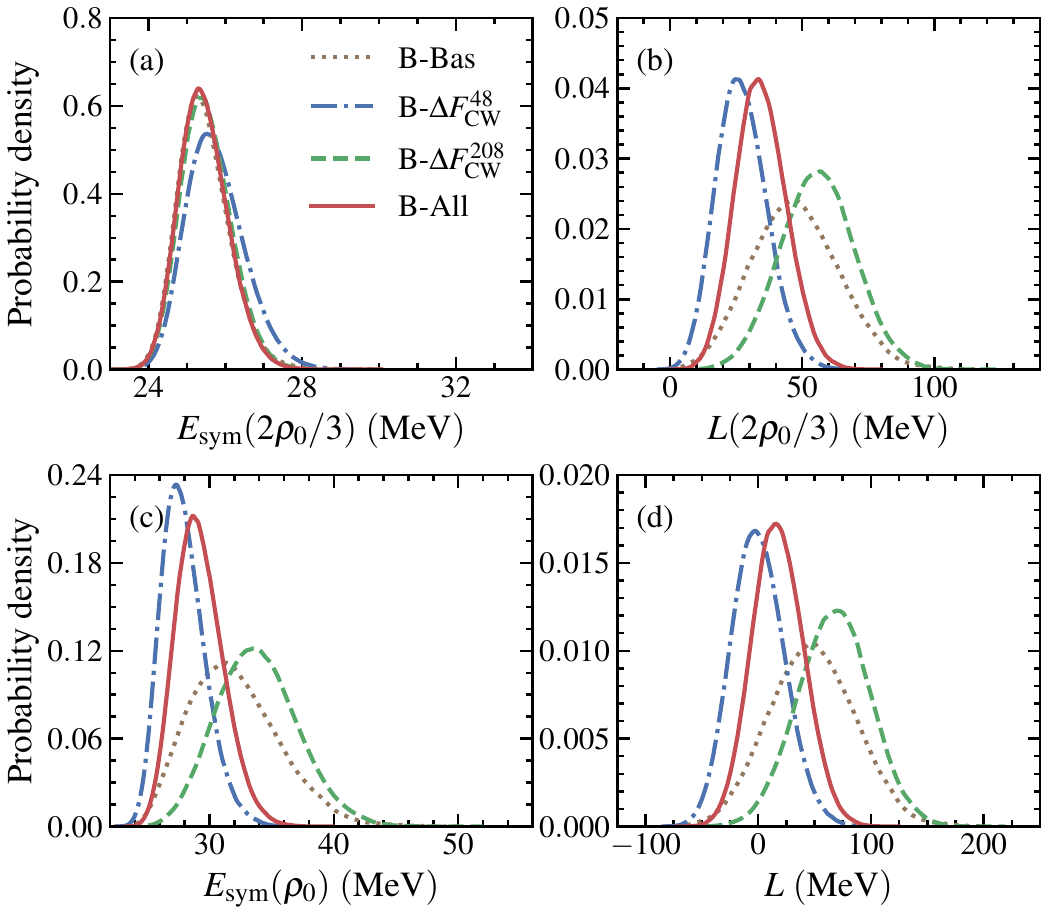}
    \caption{(Color online). Posterior distributions of $E_{\mathrm{sym}}(2\rho_0/3)$ (a),
    $L(2\rho_0/3)$ (b), $E_{\mathrm{sym}}(\rho_0)$ (c) and $L$ (d) in the four cases (see text for details). }
\label{Fig:Post}
\end{figure}

Shown in Fig.~\ref{Fig:Post} are the posterior distributions of $E_{\mathrm{sym}}(2\rho_0/3)$, $L(2\rho_0/3)$,
$E_{\mathrm{sym}}(\rho_0)$ and $L$ from the four cases.
It is seen from Fig.~\ref{Fig:Post}~(a) that for B-Bas, the $E_{\mathrm{sym}}(2\rho_0/3)$ has already been well constrained by properties of doubly magic nuclei, and further including $\Delta F_{\rm{CW}}$ has minor effects on the posterior distribution of $E_{\mathrm{sym}}(2\rho_0/3)$. With the $E_{\rm{sym}}(2\rho_0/3)$ tightly constrained, the $L(2\rho_0/3)$, $E_{\rm{sym}}(\rho_0)$ and $L$ become highly correlated, leading to very similar shapes for their posterior distributions as shown in Figs.~\ref{Fig:Post}~(b), (c) and (d).
Unlike the $E_{\rm{sym}}(2\rho_0/3)$,
the $L(2\rho_0/3)$, $E_{\mathrm{sym}}(\rho_0)$ and $L$
are all weakly constrained with B-Bas, with their posterior distributions exhibiting large distribution widths.
Comparing the results from B-$\Delta F_{\rm{CW}}^{48}$ and
B-$\Delta F_{\rm{CW}}^{208}$, one sees the tension between CREX and PREX-2.
For example, the measured $\Delta F_{\rm{CW}}^{208}$ by PREX-2  favors a larger $L$
[i.e. $68_{-33(54)}^{+32(52)}$ MeV at $68.3\%$($90\%$) C.L.], while the $\Delta F_{\rm{CW}}^{48}$ by CREX prefers a much smaller $L$ [i.e., $-1_{-23(36)}^{+24(41)}$ at $68.3\%$($90\%$) C.L.].
The $68.3\%$
credible intervals obtained from the CREX and PREX data are incompatible, whereas the $90\%$ credible intervals overlap within the region of $L = 14.2\sim40.1$ MeV. The overlap region of the $L$ distributions extracted from CREX and PREX-2 amounts for about $23\%$.

From B-All in Fig.~\ref{Fig:Post},
we find $E_{\mathrm{sym}}(2\rho_0/3)=25.4_{-0.6(0.9)}^{+0.7(1.2)}$~MeV, $L(2\rho_0/3) =34.1_{-9.2(14.8)}^{+10.1(16.8)}$~MeV, $E_{\mathrm{sym}}(\rho_0)=29.1_{-1.8(2.7)}^{+2.1(3.6)}$~MeV and $L=17.1_{-22.3(36.0)}^{+23.8(39.3)}$~MeV at $68.3\%$($90\%$) C.L..
The obtained
$E_{\rm{sym}}(2\rho_0/3)=25.4_{-0.7}^{+1.2}$ MeV at $90\%$ C.L. is consistent with $E_{\rm{sym}}(2\rho_0/3)\approx 26~\mathrm{MeV}$~\cite{Horowitz:2000xj} and $E_{\mathrm{sym}}(0.1~\mathrm{fm}^{-3})= 25.4\pm0.8~\mathrm{MeV}$~\cite{Brown2013}
obtained respectively from relativistic and nonrelativistic EDFs constrained by nuclear masses, as well as $E_{\mathrm{sym}}(0.11~\mathrm{fm}^{-3})=26.2\pm1.0$ MeV extracted from $\Delta \epsilon_F$ in doubly magic nuclei~\cite{Wang2013}
and $E_{\mathrm{sym}}(0.11~\mathrm{fm}^{-3})=26.65\pm0.2$ MeV extracted from the binding energy difference of heavy isotope pairs~\cite{Zhang:2013wna}.
The inferred
$E_{\mathrm{sym}}(\rho_0)$ and $L$ from B-All {indicate a soft symmetry energy around $\rho_0$ but are still} consistent with many previous constraints.
For example,
the upper limit of $L = 40.9$~MeV at $68.3\%$ C.L. agrees with the constraint of $L=53_{-15}^{+14}$~MeV extracted recently by combining astrophysical data, PREX-2 and chiral effective theory calculations~\cite{Essick:2021kjb}.
The inferred soft $E_{\rm sym}(\rho)$ with $L =17.1$~MeV also agrees well with the recent constraints from analyzing the $\alpha_D$ in neutron-rich Sn isotopes~\cite{Li:2021aij}.

\begin{figure}[bth]
    \centering
    \includegraphics[width=0.9\linewidth]{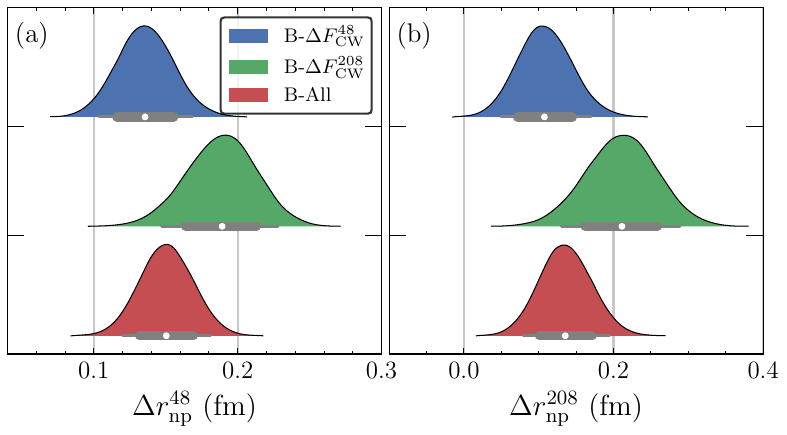}
    \caption{(Color online).  Posterior distributions of $\Delta r_{\rm{np}}^{48}$~(a) and $\Delta r_{\rm{np}}^{208}$~(b) for B-$\Delta F_{\rm{CW}}^{48}$, B-$\Delta F_{\rm{CW}}^{208}$ and B-All. Dots and bars indicate the median values, along with the $68.3\%$ and $90\%$ uncertainties}
\label{Fig:Rnp}
\end{figure}

Shown in Fig.~\ref{Fig:Rnp} are the posterior distributions of $\Delta r_{\rm{np}}^{48}$ and $\Delta r_{\rm{np}}^{208}$ for B-$\Delta F_{\rm{CW}}^{48}$, B-$\Delta F_{\rm{CW}}^{208}$ and B-All.
As expected,
the CREX(PREX-2) data result in thinner(thicker) $\Delta r_{\rm{np}}$.
Again, their $90\%$ credible intervals overlap with each other.
The extracted
$\Delta r_{\rm{np}}^{208} = 0.211_{-0.049}^{+0.047}~\mathrm{fm}$ ($68.3\%$ C.L.)
is consistent with
$0.283\pm0.071~\rm{fm}$ reported by PREX-2~\cite{PREX:2021umo}, and the extracted
$\Delta r_{\rm{np}}^{48}= 0.136_{-0.020}^{+0.020}~\mathrm{fm}$ ($68.3\%$ C.L.) also agrees well with $0.121\pm0.026(\rm{exp})\pm 0.024(\rm{model})~\rm{fm}$ by CREX~\cite{CREX:2022kgg}.
Combining the CREX and PREX data
results in $\Delta r_{\rm{np}}^{208} = 0.136_{-0.035(0.056)}^{+0.036(0.059)}$~fm and $\Delta r_{\rm{np}}^{48} = 0.150_{-0.019(0.030)}^{+0.019(0.031)}$~fm at $68.3\%$ ($90\%$) C.L..
The predicted $\Delta r_{\rm{np}}$ is relatively thin, but is still consistent with many previous experimental and theoretical studies~\cite{Zhang:2013wna,Roca-Maza:2013mla,Roca-Maza:2015eza}, e.g., the very recent {\it ab initio} predictions of $\Delta r_{\rm{np}}^{208}=0.14-0.20~\mathrm{fm}$ and $\Delta r_{\rm{np}}^{48}=0.14-0.19~\mathrm{fm}$ at $68.3\%$ C.L.~\cite{Hu:2021trw}, and the $\Delta r_{\rm{np}}^{48}=0.15-0.21$~fm from \isotope[54]{Ni}-\isotope[54]{Fe} charge radius difference~\cite{Pineda:2021shy}.
Overall,
our Bayesian analyses indicate that the CREX and PREX data are compatible with each other at $90\%$ C.L., although they are incompatible at $68.3\%$ C.L.. Furthermore, the inferred results of $E_{\mathrm{sym}}(\rho)$ and $\Delta r_{\rm{np}}$ by combining the CREX and PREX data much favor the results from CREX alone, implying the PREX-2 is less effective to constrain the $E_{\rm sym}(\rho)$ and $\Delta r_{\rm np}$ due to its lower precision of $\Delta F_{\rm CW}$ compared to the CREX.

\begin{figure}[htb]
    \centering
    \includegraphics[width=1.0\linewidth]{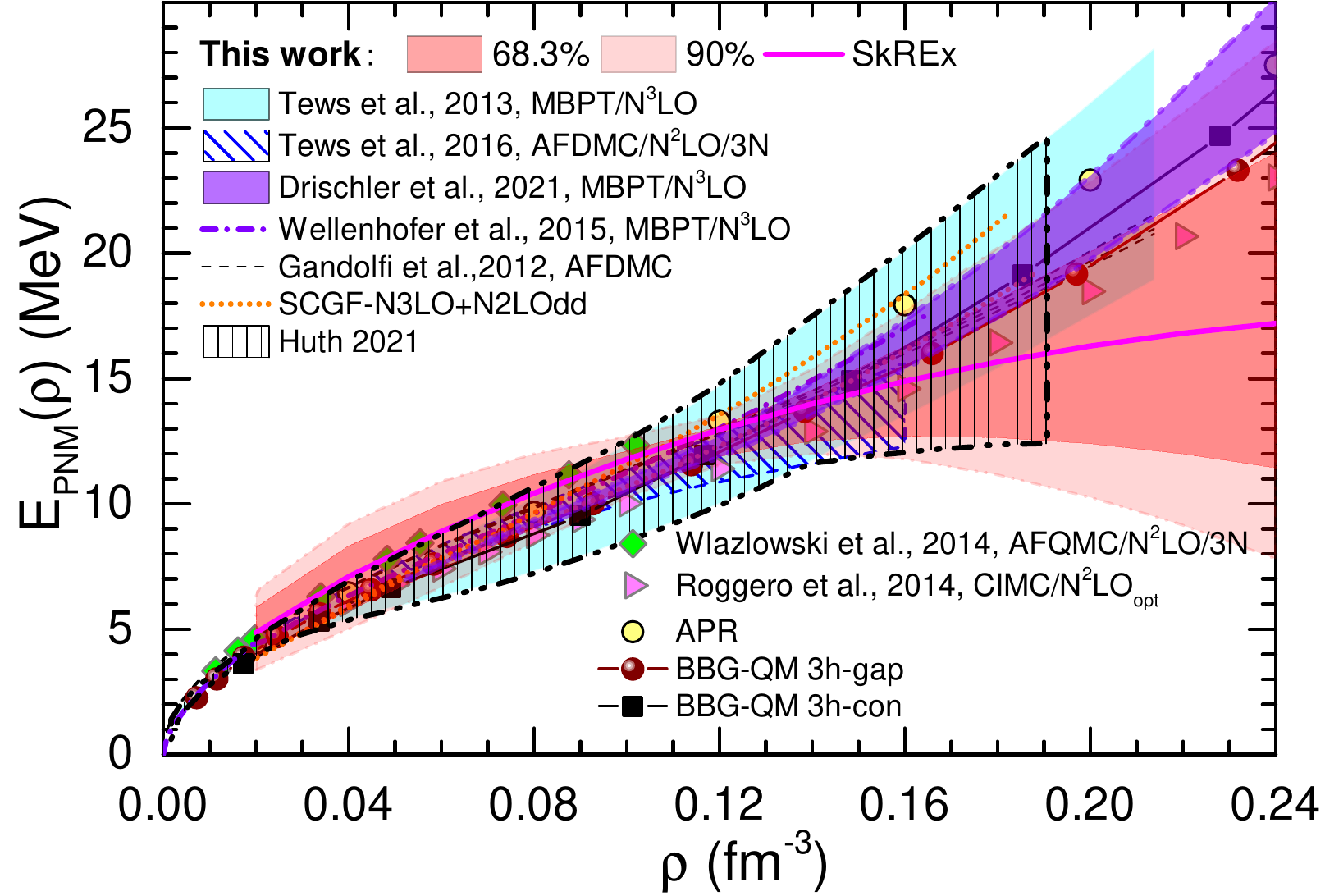}
    \caption{(Color online). The Bayesian inferred $E_{\mathrm{PNM}}(\rho)$ by combining CREX and PREX-2 data. The results from microscopic calculations and the Skyrme interaction SkREx are also included for comparison (see text for details).}
\label{Fig:Epnm}
\end{figure}

It is instructive to see the Bayesian inference on the neutron matter EOS $E_{\mathrm{PNM}}(\rho)$ which has been well constrained by microscopic calculations. Fig.~\ref{Fig:Epnm} shows the inferred $E_{\mathrm{PNM}}(\rho)$ by combining the CREX and PREX data at $68.3\%$ and $90\%$ C.L., together with the predictions
from many-body perturbation theory using N$^3$LO chiral interactions by Tews {\it et al.}~\cite{Tews:2012fj}, Wellenhofer {\it et al.}~\cite{Wellenhofer:2015qba} and Drischler {\it et al.}~\cite{Drischler:2020hwi},
the quantum Monte Carlo methods by Gandolfi {\it et al.}~\cite{Gandolfi:2011xu}, Wlaz\l{}owski {\it et al.}~\cite{Wlazlowski:2014jna},
Roggero {\it et al.}~\cite{Roggero:2014lga} and Tews {\it et al.}~\cite{Tews:2015ufa},
the variational
calculations by Akmal-Pandharipande-Ravenhall (APR)~\cite{Akmal:1998cf},
the Bethe-Bruckner-Goldstone calculations
(BBG-QM 3h-gap and BBG-QM 3h-con)~\cite{Baldo:2014rda}, and
the self-consistent Green's function
approach~(SCGF-N3LO+N2LOdd)~\cite{Carbone:2014mja}.
The region indicated by the dash-dot-dotted line in Fig.~\ref{Fig:Epnm} displays the combined constraint on the $E_{\mathrm{PNM}}(\rho)$ by various microscopic calculations (see also, Ref.~\cite{Huth:2020ozf}).
One sees that
the inferred $E_{\mathrm{PNM}}(\rho)$  agrees well with the microscopic calculations. However, at supra-saturation densities,
the inferred $E_{\mathrm{PNM}}(\rho)$ exhibits rather large uncertainties,
implying
the current data mainly constrain the $E_{\mathrm{PNM}}(\rho)$ at $\rho \lesssim \rho_0$ and
more accurate measurements on
nuclear weak form factor is necessary to effectively constrain the $E_{\mathrm{PNM}}(\rho)$
at supra-saturation densities.

\begin{figure}[htb]
  \centering
  \includegraphics[width=1.0\linewidth]{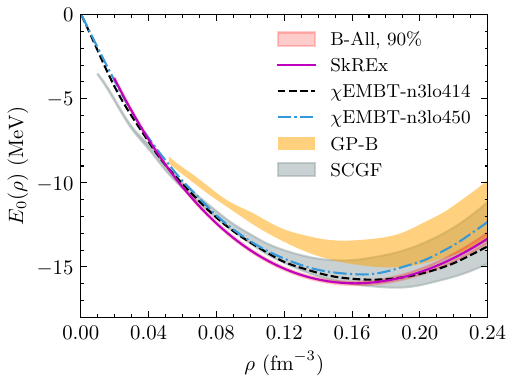}
  \caption{(Color online). The Bayesian inferred binding energy per nucleon in symmetric nuclear matter as a function of density $\rho$ at $90\%$ confidence level by combining CREX and PREX-2 data (B-All, 90\%), together with the prediction of SkREx EDF. The dashed and dash-dotted lines represent the $\chi$EMPT calculations using n3lo414 and n3lo450 forces, respectively~\cite{Wellenhofer:2015qba}.
  The orange region displays the $1\sigma$ uncertainty band derived from chiral effective theory in Ref.\cite{Drischler:2020hwi}, and the gray band is the result of SCGF approach~\cite{Xu:2019ouo}. }
\label{Fig:E0}
\end{figure}

Figure \ref{Fig:E0} exhibits the Bayesian inferred binding energy per nucleon in symmetric nuclear matter $E_0(\rho)$ as a function of nucleon density at $90\%$ confidence level by combining CREX and PREX-2 data (B-All, 90\%), together with the prediction of the SkREx EDF. For comparison, we also show  in Fig.~\ref{Fig:E0} the predictions of chiral effective many-body perturbation theory ($\chi$EMBPT) using n3lo414 and n3lo450 forces~\cite{Wellenhofer:2015qba}, the $1\sigma$ uncertainty band (GP-B) derived from chiral effective theory
using a Bayesian approach based on Gaussian process reported in Ref.\cite{Drischler:2020hwi}, and the uncertainty band of SCGF approach due to the use of three different chiral forces~\cite{Xu:2019ouo}.
One can see that, within the framework of Skyrme energy density functional, the $E_0(\rho)$ up to $1.5\rho_0$ has been well constrained by the properties of finite nuclei. On the other hand, in microscopic calculations, there are relatively large uncertainties in the predicted $E_0(\rho)$, which arise due to the choice of nuclear forces and many-body methods.

\begin{figure}[h!]
    \centering
    \includegraphics[width=0.95\linewidth]{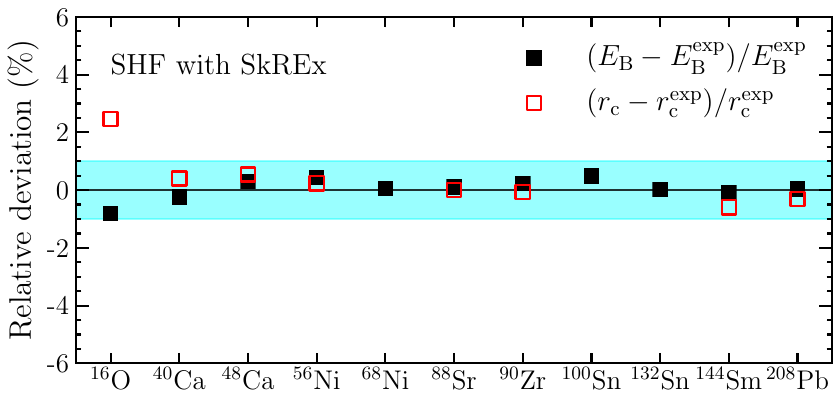}
    \caption{(Color online). Relative deviation of the binding
    energies $E_{\rm{B}}$ and charge radii $r_{\rm{c}}$ of \isotope[16]{O},
    \isotope[40]{Ca},\isotope[48]{Ca},
    \isotope[56]{Ni}, \isotope[68]{Ni},
    \isotope[88]{Sr}, \isotope[90]{Zr}, \isotope[100]{Sn},
    \isotope[132]{Sn}, \isotope[144]{Sm}, and
    \isotope[208]{Pb} obtained using the SkREx EDF
    from the experimental measurements~\cite{Wang2017,Angeli2013,Klupfel:2008af,LeBlanc:2005ik}.
    The shaded region indicates $\pm1\%$ relative deviation.\label{Fig:EbRc}}
\end{figure}

Also included in Fig.~\ref{Fig:Epnm} is the prediction from SkREx, which well agrees with the microscopic calculations.
Furthermore, Fig.~\ref{Fig:EbRc} compares the total binding energies
and charge radii of \isotope[16]{O},
\isotope[40]{Ca},\isotope[48]{Ca},
\isotope[56]{Ni}, \isotope[68]{Ni},
\isotope[88]{Sr}, \isotope[90]{Zr}, \isotope[100]{Sn},
\isotope[132]{Sn}, \isotope[144]{Sm}, and
\isotope[208]{Pb} from SkREx with the experimental
values. It is seen that the SkREx EDF overall well reproduces the experimental data with the relative deviations less than $1\%$, except for the light nucleus \isotope[16]{O} for which the mean field models are relatively less valid. 
See Table~\ref{Tab:nuclei} for
the numeric values of  SkREx  EDF predictions together with experimental data.
We also note the SkREx predicts $\Delta r_{\rm np}^{48}=0.152~\rm{fm}$, $\Delta r_{\rm{np}}^{208} = 0.141~\rm{fm}$,
$E_{\rm{sym}}(2\rho_0/3) =25.8~\rm{MeV}$ and $L(2\rho_0/3) =34.0~\rm{MeV}$.
In addition,
the $\alpha_{{D}}$ is known as an important isovector indicator~\cite{Reinhard:2010wz,Piekarewicz:2012pp,Roca-Maza:2013mla,Roca-Maza:2015eza,Zhang:2015ava,Xu:2020xib}.
Analyses based on modern nuclear EDFs suggest a strong correlation between $\alpha_{D}^{208}E_{\mathrm{sym}}(\rho_0)$ and $L$~\cite{Roca-Maza:2013mla,Roca-Maza:2015eza} as well as between $\alpha_{D}^{208}$ and $E_{\rm{sym}}(\rho_0/3)$~\cite{Zhang:2015ava,Xu:2020xib}.
The $\alpha_{D}$ in \isotope[208]{Pb} and \isotope[48]{Ca} have been determined to be
$19.6\pm0.6~\rm{fm^3}$~\cite{Tamii2011a,Roca-Maza:2015eza}  and $2.07\pm0.22~\rm{fm^3}$~\cite{Birkhan:2016qkr}, respectively, via forward-angle proton elastic scattering experiments. From CHF calculation~\cite{Reinhard:2021yke} with SkREx, we obtain $\alpha_{D}^{48} = 2.24~\rm{fm^3}$ and $\alpha_{D}^{208} = 19.5~\rm{fm^3}$, agreeing well with the data.

\begin{table}
  \caption{Experimental data~\cite{Klupfel:2008af} and predictions of SkREx energy density functional for the total binding energy $E_{\rm{B}}$ and charge radii $r_{c}$ for several typical spherical nuclei.\label{Tab:nuclei}}
  \begin{tabular}{ccccc}
  \hline \hline
  Nucleus & \multicolumn{2}{c}{$E_{\rm{B}}$ (MeV)}  & \multicolumn{2}{c}{$r_c$ (fm)} \\
 \cmidrule(lr){2-3}
  \cmidrule(lr){4-5}
& Expt. & SkREx & Expt. & SkREx \\
  \hline
 \isotope[208]{Pb} &-1636.446 & -1637.174   & 5.504&  5.487 \\
 \isotope[144]{Sm} &-1195.740  & -1194.686  & 4.960&  4.931 \\
 \isotope[132]{Sn} &-1102.900  & -1103.103 & $-$  &  4.708 \\
 \isotope[100]{Sn} &-825.800  &  -829.854   & $-$  &  4.478 \\
 \isotope[90]{Zr}  &-783.893  &  -785.757   & 4.269&  4.266 \\
 \isotope[88]{Sr}  &-768.467  &  -769.317   & 4.220&  4.220 \\
 \isotope[68]{Ni}  &-590.430  &  -590.746   & $-$&  3.889 \\
 \isotope[56]{Ni}  &-483.900  &  -485.981  & 3.750&  3.759 \\
 \isotope[48]{Ca}  &-415.990  &  -417.226   & 3.479&  3.498 \\
 \isotope[40]{Ca}  &-342.051  &  -341.227   & 3.478&  3.492 \\
 \isotope[16]{O}   &-127.620  &  -126.595   & 2.701&  2.768 \\
  \hline \hline
  \end{tabular}
\end{table}

\begin{figure}[bth]
    \centering
    \includegraphics[width=0.9\linewidth]{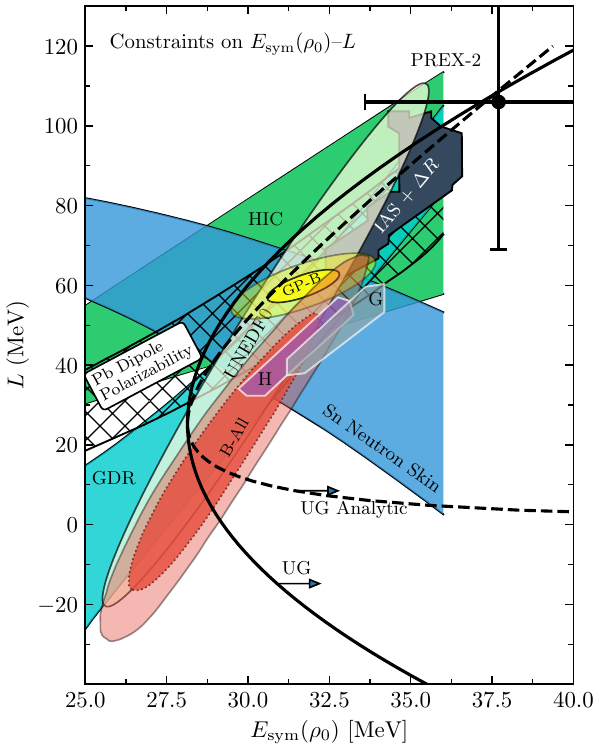}
    \caption{(Color online). Constraints on $E_{\rm{sym}}(\rho_0)$-$L$ (see text
    for details). The 68.3\% (90\%) region of the posterior joint $E_{\rm{sym}}(\rho_0)$-$L$ distribution inferred with B-All
    is shown as dark (light-red) area. Adapted from Refs.~\cite{Drischler:2020hwi,Lattimer2014}. }
\label{Fig:EsymL}
\end{figure}

Finally, we plot in Fig.~\ref{Fig:EsymL} the posterior joint $E_{\rm{sym}}(\rho_0)$-$L$ distribution in the $68.3\%$ and $90\%$ credible regions
with B-All.
For comparison, we also include the constraints summarized in Refs.~\cite{Lattimer2014, Drischler:2020hwi, Reed:2021nqk}, i.e., those from transport model analyses of mid-peripheral heavy-ion collisions (HIC)~\cite{Tsang2009b} and {$90\%$ confidence region predicted by UNEDF0 EDF}~\cite{Kortelainen2010}, the neutron skin in Sn isotopes~\cite{Chen:2010qx}, the $\alpha_D$ in \isotope[208]{Pb}~\cite{Roca-Maza:2013mla}, the centroid energy of giant dipole resonance (GDR) in \isotope[208]{Pb}~\cite{Trippa:2008gr}, the combination of isobaric analog state and isovector skins (IAS+$\Delta$R)~\cite{Danielewicz:2016bgb}, and the neutron skin in \isotope[208]{Pb} from PREX-2~\cite{PREX:2021umo, Reed:2021nqk}.
Also shown in Fig.~\ref{Fig:EsymL} are the results from microscopic calculations by Hebeler et al. (H)~\cite{Hebeler:2010jx}, Gandolfi et al. (G)~\cite{Gandolfi:2011xu} and the BUQEYE collaboration (GP-B)~\cite{Drischler:2020hwi}
as well as from the unitary gas (UG) limit by Tews et al.~\cite{Tews:2016jhi}.
Overall, the B-All suggests a soft symmetry energy, mainly due to the smaller $\Delta F_{\rm CW}$ in \isotope[48]{Ca} measured by CREX.
A soft $E_{\rm sym}(\rho)$ around $\rho_0$ will have important implications on neutron star properties.
For example, a softer $E_{\rm sym}(\rho)$ around $\rho_0$ generally gives a higher value of the neutron star core-crust transition density $\rho_t$~\cite{Xu:2009vi,Ducoin:2010as} which plays a critical role in understanding many properties of neutron stars~\cite{Pethick:1995di,Horowitz:2000xj,Oyamatsu:2006vd,Lattimer2007a}.
Using the dynamical method~\cite{Xu:2009vi},
we find that the B-All gives
$\rho_t = 0.097_{-0.016(0.023)}^{+0.026(0.049)}~\rm{fm}^{-3}$
at $68.3\%$ ($90\%$) C.L., favoring a significantly larger $\rho_t$ value
compared to the fiducial $\rho_t = 0.075~\rm{fm}^{-3}$~\cite{Link:1999ca}.
In addition,
the possible soft $E_{\rm sym}(\rho)$ at supra-saturation densities
inferred in the present work may imply that the quark-hadron phase transition may happen at a relatively low density~\cite{Zhang:2022sep}, or the non-Newtonian gravity may be needed to explain the observations of neutron stars~\cite{Wen:2009av}.
Besides its significance in neutron-star physics, 
the soft symmetry energy also has important impacts on various issues in nuclear physics studies. Notably, it has considerable effects on the location of neutron-drip line and the astrophysical r-process path~\cite{Wang:2014mra,Oyamatsu:2017qzv}, and the small $L$ value may imply the possible existence of the quasi-bound state of pure neutron matter~\cite{Oyamatsu:2017qzv}.

\section{Conclusions}
Using Bayesian inference method and the Skyrme EDF,
we have demonstrated that the CREX and PREX-2 data are compatible with each other at $90\%$ C.L., although they are incompatible at $68.3\%$ C.L..
We have further obtained a new Skyrme interaction SkREx, which can describe the CREX and PREX-2 data at $90\%$ C.L.,
the measured $\alpha_D$ in \isotope[48]{Ca} and \isotope[208]{Pb}, and the microscopic neutron matter EOS.
Our Bayesian analyses indicate that
the PREX-2 is less effective to constrain the $E_{\rm sym}(\rho)$ and $\Delta r_{\rm np}$ due to its lower precision of $\Delta F_{\rm CW}$ compared to the CREX, implying the {more precise determination of $\Delta F_{\rm CW}$ from} future {MREX} experiment~\cite{Becker:2018ggl} {or the RES-NOVA experiment via the detection of nearby core-collapse supernova neutrinos~\cite{Pattavina:2020cqc,RES-NOVA:2021gqp,Huang:2022wqu}}
is of particular importance.
Overall, the thinner neutron skin in \isotope[48]{Ca} and \isotope[208]{Pb} together with a soft $E_{\rm sym}(\rho)$ around $\rho_0$ have been inferred from combining the CREX and PREX-2 data, i.e.,
$\Delta r_{\rm{np}}^{208} = 0.136_{-0.035(0.056)}^{+0.036(0.059)}$~fm,
$\Delta r_{\rm{np}}^{48} = 0.150_{-0.019(0.030)}^{+0.019(0.031)}$~fm,
$E_{\mathrm{sym}}(\rho_0)=29.1_{-1.8(2.7)}^{+2.1(3.6)}$~MeV and
$L=17.1_{-22.3(36.0)}^{+23.8(39.3)}$~MeV at $68.3\%$($90\%$) C.L..
The soft $E_{\rm sym}(\rho)$ around $\rho_0$ will have important implications on neutron star physics and nuclear physics.

\section*{Acknowledgements}
This work was supported in part by the National Natural Science Foundation of China under Grant Nos. 12235010, 11905302 and 11625521, the National SKA Program of China No. 2020SKA0120300, and
the Fundamental Research Funds for the Central Universities, Sun Yat-Sen
University (No. 22qntd1801).

\bibliography{refCrexPrex}

\end{document}